\documentclass[preprint]{aastex}

\newcommand{\kms}   {{\rm km}\,{\rm s}^{-1}}
\newcommand{\kpc}   {{\rm kpc}}

\newcommand{\Day}   {{\rm day}}
\newcommand{\mas}   {{\rm mas}}
\newcommand{\muas}  {{\mu {\rm as}}}
\newcommand{\HJD}   {{\rm HJD}}
\newcommand{\Teff}  {T_{\rm eff}}
\newcommand{\te}    {t_{\rm E}}
\newcommand{\thetae}{\theta_{\rm E}}
\newcommand{\cc}    {{\rm cc}}
\newcommand{\phot}  {{\rm phot}}
\newcommand{\geom}  {{\rm geom}}
\newcommand{\s}     {{\rm S},0}
\newcommand{\RC}    {{\rm RC}}
\newcommand{\Dl}    {D_{\rm L}}
\newcommand{\Ds}    {D_{\rm S}}
\newcommand{\cv}    {c_{V}}
\newcommand{\ci}    {c_{I}}

\begin{document}

\title{PLANET observations of microlensing event OGLE-1999-BUL-23:\\
limb darkening measurement of the source star}

\author{
M. D. Albrow\altaffilmark{1},
J. An\altaffilmark{2},
J.-P. Beaulieu\altaffilmark{3},
J. A. R. Caldwell\altaffilmark{4},
D. L. DePoy\altaffilmark{2},\\
M. Dominik\altaffilmark{5},
B. S. Gaudi\altaffilmark{2},
A. Gould\altaffilmark{2},
J. Greenhill\altaffilmark{6},
K. Hill\altaffilmark{6},
S. Kane\altaffilmark{6},\\
R. Martin\altaffilmark{7},
J. Menzies\altaffilmark{4},
R. W. Pogge\altaffilmark{2},
K. R. Pollard\altaffilmark{8},
P. D. Sackett\altaffilmark{5},\\
K. C. Sahu\altaffilmark{1},
P. Vermaak\altaffilmark{4},
R. Watson\altaffilmark{6},
and
A. Williams\altaffilmark{7}
}
\author{(The PLANET Collaboration)}

\altaffiltext{1}{Space Telescope Science Institute,
3700 San Martin Drive, Baltimore, MD 21218, U.S.A.}
\altaffiltext{2}{Ohio State University, Department of Astronomy,
140 W 18th Avenue, Columbus, OH 43210, U.S.A.}
\altaffiltext{3}{Institut d'Astrophysique de Paris, INSU CNRS,
98 bis Boulevard Arago, F-75014, Paris, France}
\altaffiltext{4}{South African Astronomical Observatory,
P.O. Box 9, Observatory 7935, South Africa}
\altaffiltext{5}{Kapteyn Astronomical Institute,
Postbus 800, 9700 AV Groningen, The Netherlands}
\altaffiltext{6}{University of Tasmania, Physics Department,
G.P.O. 252C, Hobart, Tasmania 7001, Australia}
\altaffiltext{7}{Perth Observatory,
Walnut Road, Bickley, Western Australia 6076, Australia}
\altaffiltext{8}{Univ.\ of Canterbury, Dept.\ of Physics \& Astronomy,
Private Bag 4800, Christchurch, New Zealand}

\begin{abstract}
We present PLANET observations of OGLE-1999-BUL-23,
a binary-lens microlensing event towards the Galactic bulge.
PLANET observations in the I and V bands cover the event
from just before the first caustic crossing until
the end of the event. In particular, a densely-sampled second caustic
crossing enables us to derive the linear limb-darkening coefficients of the
source star; $\cv=0.786^{+0.080}_{-0.078}$ and $\ci=0.632^{+0.047}_{-0.037}$.
Combined analysis of the light curve and the color-magnitude diagram
suggests that the source star is a G/K subgiant in
the Galactic bulge ($\Teff\simeq 4800$ K).
The resulting linear limb-darkening coefficient
of the source is consistent with theoretical predictions,
although it is likely that non-linearity of the stellar surface brightness
profile complicates the interpretation, especially for the I band. The global
light curve fit to the data indicates that the event is due to a binary lens of
a mass ratio $q\simeq 0.39$ and a projected separation $d\simeq 2.42$.
The lens/source relative proper motion is $(22.8\pm 1.5)\ \kms\ \kpc^{-1}$,
typical of bulge/bulge or bulge/disk events.
\end{abstract}


\section{Introduction}
In point-source-point-lens (PSPL) microlensing events,
the light curve yields only one physically interesting parameter,
the characteristic time scale of the event, $\te$,
which is a combination of the mass of the lens and
the source-lens relative parallax and proper motion.
However, more varieties than PSPL events have been observed in reality,
and using deviations from the standard light curve,
one can deduce more information about the lens and the source.
The Probing Lensing Anomalies NETwork (PLANET) is an international
collaboration that monitors events in search of such anomalous
light curves using a network of telescopes in the southern hemisphere
\citep{al98}.

One example of information that can be extracted from anomalous
events is the surface brightness profile of the source star \citep{wi95}.
In a binary or multiple lens system, the caustic is an extended structure.
If the source passes near or across the caustic,
drastic changes in magnification near the caustics
can reveal the finite size of the source \citep{go94, ne94, wi94, al97},
and one can even extract its surface-brightness profile
\citep{bc96, go96, sa97, va98}.

The fall-off of the surface brightness near the edge of the stellar disk
with respect to its center, known as limb darkening,
has been extensively observed in the Sun.
Theories of stellar atmospheres predict limb darkening as a general phenomenon
and give models for different types of stars.
Therefore, measurement of limb darkening in distant stars other than the Sun
would provide important observational constraints
on the study of stellar atmospheres.
However, such measurements are very challenging with traditional techniques
and have usually been restricted to relatively nearby stars
or extremely large supergiants. As a result, only a few attempts have 
been made to measure limb darkening to date. The classical method
of tracing the stellar surface brightness profile is the analysis of the light
curves of eclipsing binaries \citep{wi71, tw80}.
However, the current practice in eclipsing-binary studies
usually takes the opposite approach to limb darkening
\citep{cl98a} -- constructing models of light
curves using theoretical predictions of limb darkening.
This came to dominate after \citet{po84}
demonstrated that the uncertainty of limb darkening measurements
from eclipsing binaries is substantially larger than the theoretical
uncertainty. Since the limb-darkening parameter is highly correlated
with other parameters of the eclipsing binary,
fitting for limb darkening could seriously degrade the
measurement of these other parameters.
Multi-aperture interferometry and lunar occultation,
which began as measurements of the angular sizes of stars,
have also been used to resolve the surface structures of stars \citep{ho98}.
In particular, a large wavelength dependence of the interferometric size
of a stellar disk has been attributed to limb darkening,
and higher order corrections to account for limb darkening
have been widely adopted in the interferometric
angular size measurement of stars.
Several recent investigations using optical interferometry
extending beyond the first null of the visibility function have indeed
confirmed that the observed patterns of the visibility function
contradict a uniform stellar disk model
and favor a limb-darkened disk \citep{qu96, ha98}
although these investigations have used
a model prediction of limb darkening
inferred from the surface temperature rather than
deriving the limb darkening from the observations.
However, at least in one case, \citet{bu97} used interferometric imaging to
measure the stellar surface brightness profile with coefficients beyond the
simple linear model.
In addition, developments of high resolution
direct imaging in the last decade using 
space telescopes \citep{gi96} or speckle imaging \citep{kl97}
have provided a more straightforward way of detecting
stellar surface irregularities. However, most studies of this kind are still
limited to a few extremely large supergiants, such as $\alpha$ Ori.
Furthermore, they seem to be more sensitive to asymmetric surface structures
such as spotting than to limb darkening.

By contrast, microlensing can produce limb-darkening measurements
for distant stars with reasonable accuracy.
To date, limb darkening (more precisely, a set of coefficients of 
a parametrized limb-darkened profile) has
been measured for source stars in three events, two K giants in the Galactic
bulge and an A dwarf in the Small Magellanic Cloud (SMC). MACHO 97-BLG-28
was a cusp-crossing event of a K giant source with extremely good data,
permitting \citet{al99a} to make a two-coefficient
(linear and square-root) measurement of limb darkening.
\citet{af00} used data
from five microlensing collaborations to
measure linear limb darkening coefficients in five filter bandpasses for
MACHO 98-SMC-1, a metal-poor A star in the SMC. Although the data for this
event were also excellent, the measurement did not yield a two-parameter
determination because the caustic crossing was a fold-caustic rather than
a cusp, and these are less sensitive to the form of the stellar surface
brightness profile.
\citet{al00a} measured a linear limb-darkening
coefficient for MACHO 97-BLG-41,
a complex rotating-binary event with both a cusp crossing and a
fold-caustic crossing. In principle, such an event could give very detailed
information about the surface brightness
profile. However, neither the cusp nor the fold-caustic crossing was
densely sampled, so only a linear parameter could be extracted.

In this paper,
we report a new limb-darkening measurement of a star in the Galactic bulge
by a fold-caustic crossing event, OGLE-1999-BUL-23,
based on the photometric monitoring of PLANET.

\section{OGLE-1999-BUL-23}
OGLE-1999-BUL-23 was originally discovered
towards the Galactic bulge by the Optical Gravitational Lensing
Experiment (OGLE) \footnote{The OGLE alert for this event is posted at
http://www.astrouw.edu.pl/\~{ }ftp/ogle/ogle2/ews/bul-23.html}
\citep{ud92, ud97}.
The PLANET collaboration observed the event as a part of our routine monitoring
program after the initial alert, and detected a
sudden increase in brightness on 12 June 1999.\footnote{
The PLANET anomaly and caustic alerts are found at
http://www.astro.rug.nl/\~{ }planet/OB99023cc.html}
Following this
anomalous behavior, we began dense (typically one observation
per hour) photometric sampling of the event.
Since the source lies close to the (northern) winter solstice
($\alpha = 18^{h} 07^{m} 45\fs14$, $\delta = -27\arcdeg 33\arcmin 15\farcs4$),
while the caustic crossing occurred nearly at the summer solstice
(19 June 1999), and since good weather at all four
of our southern sites prevailed throughout,
we were able to obtain nearly continuous coverage of
the second caustic crossing without any significant gaps.
Visual inspection and initial analysis of the light curve
revealed that the second crossing was due to a simple fold
caustic crossing (see \S 2.2).

\subsection{Data}
We observed OGLE-1999-BUL-23 with I and V band filters
at four participant telescopes:
the Elizabeth 1 m at South African Astronomical Observatory (SAAO),
Sutherland, South Africa;
the Perth/Lowell 0.6 m telescope at Perth, Western Australia;
the Canopus 1 m near Hobart, Tasmania, Australia;
and the Yale/AURA/Lisbon/OSU 1 m at Cerro Tololo
Inter-American Observatory (CTIO), La Serena, Chile.
From June to August 1999 ($1338<\HJD'<1405$),
PLANET obtained almost 600 images of the field of OGLE-1999-BUL-23. 
In addition, baseline points were taken at SAAO ($\HJD'\simeq 1440$)
and Perth ($\HJD'\simeq 1450$; $\HJD'\simeq 1470$). 
Here $\HJD'\equiv\HJD-2450000$,
where HJD is Heliocentric Julian Date at center of exposure.
The data reduction and photometric measurements of the event were
performed relative to non-variable stars in the same field using DoPHOT.
After several re-reductions, we recovered the photometric measurements
from a total of 476 frames.

We assumed independent photometric systems for different observatories
and thus explicitly included the determination
of independent (unlensed) source and
background fluxes for each different telescope and
filter band in the analysis.
This provides both determinations of the
photometric offsets between different systems
and independent estimates of the blending factors.
The final results demonstrate satisfactory alignment
among the data sets (see \S 2.3),
and we therefore believe that we have reasonable relative calibrations.
Our previous studies have shown that the background flux
(or blending factors) may correlate with the size of seeing disks
in some cases \citep{al00a, al00b}.
To check this, we introduced
linear seeing corrections in addition to constant backgrounds.

From previous experience, it is expected that the formal errors reported
by DoPHOT underestimate the actual errors \citep{al98},
and consequently that $\chi^{2}$ is overestimated. Hence,
we renormalize photometric errors to force the final reduced
$\chi^{2}/dof = 1$ for our best fit model. Here, $dof$ is the number of
degrees of freedom (the number of data points less the number of parameters).
We determine independent rescaling factors
for the photometric uncertainties from the different
observatories and filters. The process involves two steps: the elimination
of bad data points and the determination of error normalization factors.
In this as in all previous events that we have analyzed,
there are outliers discrepant by many $\sigma$ that cannot be attributed to
any specific cause even after we eliminate some points whose source of 
discrepancy is identifiable. Although,
in principle, whether particular data points are faulty or not should be 
determined without any reference to models, we find that
the light curves of various models that yield reasonably good fits to the data
are very similar to one another, and furthermore,
there is no indication of temporal clumping of highly discrepant
points. We, therefore, identify outlier points
with respect to our best model and exclude them from the final analysis.

For the determination of outliers, we follow an iterative approach using both
steps of error normalization. First, we calculate the individual
$\chi^{2}$'s of data sets from different observatories and filter bands 
with reference to our best model without any rejection or error scaling.
Then, the initial normalization factors
are determined independently for each data set
using those individual $\chi^{2}$'s and the number of data points in each set.
If the deviation of the most discrepant outlier is larger than
what is predicted based on the number of points and the assumption 
of a normal distribution, we classify the point
as bad and calculate the new $\chi^{2}$'s and the
normalization factors again.
We repeat this procedure until the largest outlier is comparable
with the prediction of a normal distribution.
Although the procedure appears somewhat arbitrary,
the actual result indicates that
there exist rather large decreases of $\sigma$
between the last rejected and included data points.
After rejection of bad points,
429 points remain (see Table~1 and Fig.~1).

\subsection{Analysis: searching for $\chi^{2}$ minima}
We use the method of \citet[hereafter Paper~I]{al99b},
which was devised to fit the light curve of fold-caustic crossing
binary-lens events, to analyze the light curve of this event and
find an appropriate binary-lens solution.
This method consists of three steps: (1) fitting of caustic-crossing
data using an analytic approximation of the magnification,
(2) searching for $\chi^{2}$ minima over the whole parameter space
using the point-source approximation and 
restricted to the non-caustic crossing data,
and (3) $\chi^{2}$ minimization using all data and 
the full binary-lens equation
in the neighborhood of the minima found in the second step.

For the first step, we fit the I-band caustic crossing data 
($1348.5 \leq \HJD' \leq 1350$) to the six-parameter analytic
curve shown in equation (1) that characterizes the shape of the
second caustic crossing (Paper~I; Afonso et al.\ 2000),
\begin{mathletters}
\begin{equation}
F(t)=\left(\frac{Q}{\Delta t}\right)^{1/2}
\left[G_{0}\left(\frac{t-t_{\cc}}{\Delta t}\right)
+\Gamma H_{1/2}\left(\frac{t-t_{\cc}}{\Delta t}\right)\right]+F_{\cc}
+(t-t_{\cc})\,\tilde{\omega}\,,
\end{equation}
\begin{equation}
G_{n}(\eta)\equiv
\pi^{-1/2}
\frac{(n+1)!}{(n+1/2)!}
\int_{\max(\eta,-1)}^{1} dx\,
\frac{(1-x^{2})^{n+1/2}}{(x-\eta)^{1/2}}\,
\Theta(1-\eta)\,,
\end{equation}
\begin{equation}
H_{1/2}(\eta)\equiv G_{1/2}(\eta)-G_{0}(\eta)\,.
\end{equation}
\end{mathletters}
Figure~2 shows the best-fit curve and the data points used for the fit.
This caustic-crossing fit essentially constrains the search for a full
solution to a four-dimensional hypersurface instead of the whole
nine-dimensional parameter space (Paper~I). 

We construct a grid of point-source light curves with model
parameters spanning a large subset of the hypersurface
and calculate $\chi^{2}$ for each model
using the I-band non-caustic crossing data.
After an extensive search for $\chi^{2}$-minima over the 
four-dimensional hypersurface, we find positions of
two apparent local minima, each in a local valley of the
$\chi^{2}$-surface.
The smaller $\chi^{2}$ of the two
is found at $(d,q,\alpha) \simeq (2.4,\,0.4,\,75\degr)$,
where $d$ is the projected binary separation in
units of the Einstein ring radius,
$q$ is the mass ratio of the binary system, and $\alpha$ is 
the angle between the binary axis and the path of the source,
defined so that the geometric center of the lens system lies on
the right hand side of the moving source.
The other local minimum is
$(d,q,\alpha) \simeq (0.55,\,0.55,\,260\degr)$.
The results appear to suggest a rough symmetry of
$d \leftrightarrow d^{-1}$ and
$(\alpha < \pi) \leftrightarrow (\alpha > \pi)$,
as was found for MACHO 98-SMC-1 (Paper I; Afonso et al.\ 2000).
Besides these two local minima,
there are several isolated $(d,q)$-grid points at which
$\chi^{2}$ is smaller than at neighboring grid points. However, on a finer grid
they appear to be connected with one of the two local minima specified above.
We include the two local minima and some of the apparently isolated minimum
points as well as points
in the local valley around the
minima as starting points for the refined search of
$\chi^{2}$-minimization in the next step.

\subsection{Solutions: $\chi^{2}$ minimization}
Starting from the local minima found in \S 2.2 and the points
in the local valleys around them,
we perform a refined search for the $\chi^{2}$ minimum.
The $\chi^{2}$ minimization includes all the I and V data points
for successive fitting to the full expression for magnification,
accounting for effects of a finite source size and limb darkening.

As described in Paper~I, the third step makes use of a variant of
equation (1) to evaluate the magnified flux in the neighborhood of
the caustic crossing. Paper~I found that, for MACHO 98-SMC-1,
this analytic expression was an extremely good approximation to
the results of numerical integration and assumed that the same would
be the case for any fold crossing. Unfortunately, we find that,
for OGLE-1999-BUL-23, this approximation deviates from the true
magnification as determined using the method of \citet{go97} as much as 4\%,
which is larger than our typical photometric uncertainty in the region of
caustic crossing.
To maintain the computational efficiency of Paper~I,
we continue to use the analytic formula~(1), but correct it by
pre-tabulated amounts given by the fractional difference (evaluated
close to the best solution) between this approximation and the values
found by numerical integration. We find that this correction works quite
well even at the local minimum for the other (close-binary) solution --
the error is smaller than 1\%, and in particular, the calculations agree
within 0.2\% for the region of primary interest. The typical (median)
photometric uncertainties for the same region are 0.015 mag (Canopus
after the error normalization) and 0.020 mag (Perth).
In addition, we test the correction by running the fitting program
with the exact calculation at the minimum found using the corrected
approximation, and find that the measured parameters change less
than the precision of the measurement. In particular, the limb-darkening
coefficients change by an order of magnitude less than the
measurement uncertainty due to the photometric errors.

The results of the refined $\chi^{2}$ minimization are listed in Table~2
for three discrete ``solutions'' and in Table~3 for grid points neighboring
the best-fit solution whose $\Delta \chi^{2}$ is less than one.
The first seven columns describe seven of the standard parameters of the
binary-lens model (the remaining two parameters are 
the source and background flux).
The eighth column is the time of the second caustic crossing
($t_{\cc}$) -- the time when the center of the source crossed the caustic.
The limb darkening coefficients for I and V bands are shown in the next
two columns. The final column is $\Delta \chi^{2}$,
\begin{equation}
\Delta \chi^{2} \equiv
\frac{\chi^{2}-\chi^{2}_{{\rm best}}}{\chi^{2}_{{\rm best}}/dof}\ ,
\end{equation}
as in Paper~I.
The light curve (in magnification) of the best-fit model is shown
in Figure~3 together with all the data points
used in the analysis.

\subsubsection{``Degeneracy''}
For typical binary-lens microlensing events,
more than one solution often fits the observations reasonably well.
In particular, \citet{do99} predicted a degeneracy between close and wide
binary lenses resulting from a symmetry in the lens equation itself,
and such a degeneracy was found empirically for MACHO 98-SMC-1
(Paper~I; Afonso et al.\ 2000).

We also find two distinct local $\chi^{2}$ minima (\S 2.2) that appear
to be closely related to such degeneracies.
However, in contrast to the case of MACHO 98-SMC-1,
our close-binary model for OGLE-1999-BUL-23 has
substantially higher $\chi^{2}$ than the wide-binary model
($\Delta\chi^{2}=\,127.86$). 
Figure~4 shows the predicted light curves in SAAO instrumental I band.
The overall geometries of these two models are shown in Figures~5 and 6.
The similar morphologies of the caustics with respect to the path
of the source is responsible for the degenerate light curves near the
caustic crossing (Fig.~6). However, the close-binary model 
requires a higher blending fraction and lower baseline flux than
the wide-binary solution because
the former displays a higher peak magnification
($A_{\max}\sim 50$ vs.\ $A_{\max}\sim 30$). 
Consequently, a precise determination of the baseline can significantly
contribute to discrimination between the two models,
and in fact, the actual data did constrain the baseline well enough 
to produce a large difference in $\chi^{2}$.

A fair number of pre-event baseline measurements
are available via OGLE, and those data can further help discriminate
between these two ``degenerate'' models. We fit OGLE measurements
to the two models with all the model parameters being fixed and
allowing only the baseline and the blending fraction as free parameters.
We find that the PLANET wide-binary model produces $\chi^{2}=306.83$
for 169 OGLE points ($\chi^{2}/dof=1.83$, compare Table~1)
while $\chi^{2}=608.22$ for the close-binary model
for the same 169 points (Fig.~7). That is, $\Delta \chi^{2}=164.04$,
so that the addition of OGLE data by itself discriminates between
the two models approximately as well as all the PLANET data combined.
The largest contribution to this large $\Delta \chi^{2}$ appears to
come from the period about a month before the first caustic crossing
which is well covered by the OGLE data but not by the PLANET data.
In particular,
the close-binary model predicts a bump in the light curve around
$\HJD'\simeq 1290$ due to a triangular caustic (see Fig.~5), but
the data do not show any abnormal feature in the same region,
although it is possible that rotation of the binary moved the
caustic far from the source trajectory (eg.\ Afonso et al.\ 2000).
In brief, the OGLE data strongly favor the wide-binary model.

\subsubsection{Limb-Darkening Coefficients}
The limb darkening of the source is parametrized
using a normalized linear model of the source surface brightness profile,
which was introduced in Appendix B of Paper I,
\begin{eqnarray}
S_{\lambda}(\vartheta)
=\bar{S}_{\lambda}
\left[1-\Gamma_{\lambda}\left(1-{3\over2}\,\cos\vartheta\right)\right]
=\bar{S}_{\lambda}
\left[(1-\Gamma_{\lambda})+{3\over2}\,\Gamma_{\lambda}\cos\vartheta\right]
\ ,\nonumber\\
{\rm where}\ \sin\vartheta\equiv\frac{\theta}{\theta_{\ast}}
\ {\rm and}\ \bar{S}_{\lambda}
\equiv\frac{F_{s,\lambda}}{\pi \theta_{\ast}^{2}}\ ,
\end{eqnarray}
while linear limb darkening is usually parametrized by,
\begin{equation}
S_{\lambda}(\vartheta)
=S_{\lambda}(0)\left[1-c_{\lambda}\,(1-\cos\vartheta)\right]\\
=S_{\lambda}(0)\left[(1-c_{\lambda})+c_{\lambda}\,\cos\vartheta\,\right]\ .
\end{equation}
The relationship between the two expressions of linear limb-darkening 
coefficients is
\begin{equation}
c_{\lambda}
=\frac{3\Gamma_{\lambda}}{2+\Gamma_{\lambda}}\ .
\end{equation}

Amongst our six data sets,
data from SAAO did not contain points that were
affected by limb darkening,
i.e.\ caustic crossing points. Since the filters used at different
PLANET observatories do not
differ significantly from one another, we use the same
limb-darkening coefficient for the three remaining I-band data sets.
The V-band coefficient is determined only from Canopus data,
so that a single coefficient is used automatically.

For the best-fit lens geometry, the measured values of linear
limb-darkening coefficients are
$\Gamma_{I}=\,0.534\pm 0.020$ and $\Gamma_{V}=\,0.711\pm 0.089$,
where the errors include only uncertainties in the linear
fit due to the photometric
uncertainties at \emph{fixed} binary-lens model parameters.
However, these errors underestimate the actual uncertainties of the
measurements because the measurements are correlated with the determination
of the seven lens parameters shown in Tables~2 and~3.
Incorporating these additional uncertainties in the
measurement (see the next section for a detailed discussion of the error
determination), our final estimates are
\begin{mathletters}
\begin{equation}
\Gamma_{I}=\,0.534^{+0.050}_{-0.040}\qquad
\left(\ci=\,0.632^{+0.047}_{-0.037}\right),
\end{equation}
\begin{equation}
\Gamma_{V}=\,0.711^{+0.098}_{-0.095}\qquad
\left(\cv=\,0.786^{+0.080}_{-0.078}\right).
\end{equation}
\end{mathletters}
This is consistent with the result of the caustic-crossing fit of \S 2.2
($\Gamma_{I}=\,0.519\pm 0.043$). Our result suggests that the source
is more limb-darkened in $V$ than in $I$, which is generally expected
by theories. Figure~8 shows the I-band residuals (in magnitudes) at the
second caustic crossing from our best-fit models for a linearly
limb-darkened and a uniform disk model. It is clear that the uniform disk
model exhibits larger systematic residuals near the peak than the
linearly limb-darkened disk. From the residual patterns -- the uniform
disk model produces a shallower slope for
the most of the falling side of the second
caustic crossing than the data require, one can infer that the source
should be more centrally concentrated than the model predicts, and
consequently the presence of limb darkening. The linearly limb-darkened
disk reduces the systematic residuals by a factor of $\sim$ 5.
Formally,
the difference of $\chi^{2}$ between the two models is 172.8 with two
additional parameters for the limb-darkened disk model, i.e.\ the
data favor a limb-darkened disk
over a uniform disk at very high confidence.

\section{Error Estimation for Limb Darkening Coefficients}
Due to the multi-parameter character of the fit,
a measurement of any parameter is correlated with other parameters
of the model. The limb-darkening coefficients obtained with
the different model parameters shown in Table~3 exhibit a considerable
scatter, and in particular, for the I-band measurement,
the scatter is larger than the uncertainties due to the photometric errors.
This indicates that, in the measurement of the limb-darkening coefficients,
we need to examine errors that correlate with
the lens model parameters in addition to the uncertainties resulting from
the photometric uncertainties at fixed lens parameters.
This conclusion is reinforced by the fact that the error in the estimate
of $\Gamma$ from the caustic-crossing fit (see Fig.~2), which includes
the correlation with the parameters of the caustic-crossing, is
substantially larger than the error in the linear fit, which does not.

Since limb darkening manifests itself mainly around the caustic crossing,
its measurement is most strongly correlated with $\Delta t$ and $t_{\cc}$.
To estimate the effects of these correlations,
we fit the data to models with $\Delta t$ or $t_{\cc}$
fixed at several values near the best fit -- the global geometry of the 
best fit, i.e.\ $d$ and $q$ being held fixed as well.
The resulting distributions of $\Delta \chi^{2}$ have parabolic shapes
as a function of the fit values of the limb-darkening coefficient
and are centered at the measurement of the best fit.
(Both, $\Delta t$ fixed and $t_{\cc}$ fixed, produce essentially the same
parabola, and therefore, we believe that the uncertainty related to
each correlation with either $\Delta t$ or $t_{\cc}$ is, in fact, same
in its nature.)
We interpret the half width of the parabola at $\Delta \chi^{2}=1$
($\delta \Gamma_{I}=0.031,\ \delta \Gamma_{V}=0.032$) as the uncertainty due to
the correlation with the caustic-crossing parameters at a given global lens
geometry of a fixed $d$ and $q$.

Although the global lens geometry should not \emph{directly} affect the
limb darkening measurement, the overall correlation between
local and global parameters can contribute an additional
uncertainty to the measurement. This turns out to be
the dominant source of the scatter found in Table~3. To incorporate this
into our final determination of errors, we examine the varying range of
the measured coefficients over $\Delta \chi^{2} \leq 1$. The result is
apparently asymmetric between the direction of increasing and decreasing
the amounts of limb darkening. We believe that this is real, and thus
we report asymmetric error bars for the limb-darkening measurements.

The final errors of the measurements reported in \S 2.3.2 are determined
by adding these two sources of error to the photometric uncertainty
in quadrature. The dominant
source of errors in the I-band coefficient measurement is the correlation
between the global geometry and the local parameters whereas the photometric
uncertainty is the largest contribution to the uncertainties in the V-band
coefficient measurement.

Although the measurements of V and I band limb darkening at fixed model
parameters are independent, the final estimates of two
coefficients are not actually independent for the same reason discussed above.
(The correlation between V and I limb-darkening coefficients is clearly
demonstrated in Table~3.) Hence, the complete description of the uncertainty
requires a covariance matrix.
\begin{mathletters}
\begin{equation}
C=
C_{\phot}\ +\ 
\tilde{C}_{\cc}^{1/2}
\left(\begin{array}{cc}
1&\xi\\
\xi&1
\end{array}\right)
\tilde{C}_{\cc}^{1/2}\ +\
\tilde{C}_{\geom}^{1/2}
\left(\begin{array}{cc}
1&\xi\\
\xi&1
\end{array}\right)
\tilde{C}_{\geom}^{1/2}\ ,
\end{equation}
\begin{equation}
C_{\phot}\equiv
\left(\begin{array}{cc}
\sigma_{V,{\rm phot}}^{2}&0\\
0&\sigma_{I,{\rm phot}}^{2}
\end{array}\right)\ ,
\end{equation}
\begin{equation}
\tilde{C}_{\cc}^{1/2}\equiv
\left(\begin{array}{cc}
\sigma_{V,\cc}&0\\
0&\sigma_{I,\cc}
\end{array}\right)\ ,
\end{equation}
\begin{equation}
\tilde{C}_{\geom}^{1/2}\equiv
\left(\begin{array}{cc}
\bar{\sigma}_{V,\geom}&0\\
0&\bar{\sigma}_{I,\geom}
\end{array}\right)\ ,
\end{equation}
\end{mathletters}
where the subscript (phot) denotes the uncertainties due to the photometric
errors; (cc), the correlation with $\Delta t$ and $t_{\cc}$ at a fixed $d$
and $q$; (geom), the correlation with the global geometry, and $\xi$ is the
correlation coefficient between $\Gamma_{V}$ and $\Gamma_{I}$ measurement.
We derive the correlation coefficient using each measurement of $\Gamma_{V}$
and $\Gamma_{I}$, and the result indicates that two measurements are almost
perfectly correlated $(\xi=0.995)$. We accommodate asymmetry of the errors
by making the error ellipse off-centered with respect to the best estimate.
(See \S 5 for more discussion on the error ellipses.)

\section{Physical Properties of the Source Star}
Figure~9 shows color-magnitude diagrams (CMDs)
derived from a $2'\times 2'$ SAAO field
and a $4'\times 4'$ Canopus field centered on OGLE-1999-BUL-23
with positions marked for the unmagnified source (S), 
the baseline (B), blended light (BL) at median seeing, and the center of
red clump giants (RC). The source position in these CMDs is consistent
with a late G or early K subgiant in the Galactic bulge (see below).
Using the color and magnitude of red clump giants
in the Galactic bulge reported by \citet{pa99}
($I_{\RC}=\,14.37 \pm 0.02,\ [V-I]_{\RC}=\,1.114 \pm 0.003$),
we measure the reddening-corrected color and magnitude of the source
in the Johnson-Cousins system from the relative position of the source
with respect to the center of red clump in our CMDs, and obtain:
\begin{mathletters}
\begin{equation}
(V-I)_{\s}=\,1.021\pm 0.044,
\end{equation}
\begin{equation}
V_{\s}=\,18.00\pm 0.06,
\end{equation}
\end{mathletters}
where the errors include the difference of the source positions
in the two CMDs, but may still be somewhat underestimated
because the uncertainty in the selection of red clump giants in our CMDs
has not been quantified exactly.

From this information, we derive the surface temperature of the source;
$\Teff=\,(4830\pm 100)$ K, using the color calibration in \citet{be98}
and assuming $\log\,g=\,3.5$ and the solar abundance.
This estimate of temperature is only weakly dependent on the assumed
surface gravity and different stellar atmospheric models.
To determine the angular size of
the source, we use equation (4) of \citet{al00a}, which is derived
from the surface brightness-color relation of \citet{vb99}.
We first convert $(V-I)_{\s}$ into $(V-K)_{\s}=\,2.298\pm 0.113$
using the same color calibration of Bessell et al.\ (1998) and then obtain
the angular radius of the source of
\begin{eqnarray}
\theta_{\ast}&=&(1.86\pm 0.13)\ \muas\nonumber\\
&=&(0.401\pm 0.027)\ R_{\sun}\ \kpc^{-1}.
\end{eqnarray}
If the source is at the Galactocentric distance (8 kpc),
this implies that the radius of the source
is roughly 3.2$R_{\sun}$, which is consistent with the size
of a $\sim 1 M_{\sun}$ subgiant ($\log\,g=\,3.4$).

Combining this result with the parameters of the best-fit model yields
\begin{eqnarray}
\mu=\theta_{\ast}/(\Delta t\ \sin\,\phi)
&=&(13.2\pm 0.9)\ \muas\ \Day^{-1}\nonumber\\
&=&(22.8\pm 1.5)\ \kms\ \kpc^{-1},\\
\thetae=\mu\,\te&=&(0.634 \pm 0.043)\ \mas,
\end{eqnarray}
where $\phi=\,123\fdg 9$ is the angle that the source crosses
the caustic (see Fig.~6).
This corresponds to a projected relative velocity of $(182\pm 12)\,\kms$
at the Galactocentric distance, which is generally
consistent with what is expected in typical bulge/bulge
or bulge/disk (source/lens) events, but inconsistent
with disk/disk lensing. Hence we conclude that the source is in the bulge.
As for properties of the lens,
the projected separation of the binary lens is
$(1.53\pm 0.10)$ AU $\kpc^{-1}$, and the combined mass of the lens is
given by
\begin{equation}
M_{{\rm L}}=\frac{c^{2} \Ds \Dl}{4G(\Ds-\Dl)}\,\thetae^{2}=
(0.395\pm 0.053)\left(\frac{x}{1-x}\right)
\left(\frac{\Ds}{8\,\kpc}\right)\,M_{\sun}\ ,
\end{equation}
where $x\equiv\,\Dl/\Ds$, $\Dl$ is the distance to the lens,
and $\Ds$ is the distance to the source.

\section{Limb Darkening of the Source}
We compare our determination of the linear limb-darkening coefficients to
model calculations by Claret, D\'{\i}az-Cordov\'{e}s, \& Gim\'{e}nez (1995)
and D\'{\i}az-Cordov\'{e}s, Claret, \& Gim\'{e}nez (1995).
For an effective temperature of $\Teff=\,(4830\pm
100)$ K and a surface gravity of $\log\,g=\,3.5$, the interpolation of 
the V-band linear limb-darkening coefficients, $c_{V}$, of \citet{di95}
predicts a value $\cv=\,0.790\pm 0.012$, very consistent
with our measurement. However, for the I-band coefficient, the prediction
of \citet{cl95}, $\ci=\,0.578\pm 0.008$, is only marginally
consistent with our measurement, at the $1.46\,\sigma$ level. Adopting a
slightly different gravity does not qualitatively change this general result.
Since we believe that the uncertainty in the color of the source is larger
than in the limb-darkening coefficients, we also examine the opposite
approach to the theoretical calculations $-$ using the measured values of
limb-darkening coefficients to derive the effective temperature of the source.
If the source is a subgiant ($\log\,g\simeq\,3.5$) as our CMDs suggest, the
measured values of the limb-darkening coefficients are expected to be observed
in stars of the effective temperature, $\Teff=\,(4850^{+650}_{-670})$ K
for $\cv$ or $\Teff=\,(4200^{+390}_{-490})$ K for $\ci$. As before, the
estimate from the V-band measurement shows a better agreement
with the measured color than the estimate from the I-band.
Considering that the data quality of I band is better than V band (the
estimated uncertainty is smaller in $I$ than in $V$), this result
needs to be explained.

In Figure~10, we plot theoretical calculations of $(\ci,\,\cv)$ together with
our measured values. In addition to \citet{di95} and \citet{cl95} (A), we
also include the calculations of linear limb-darkening coefficients by
\citet{vh93} (B) and \citet{cl98b} (C). For all three calculations, the V-band
linear coefficients are generally consistent with the measured coefficients
and the color, although \citet{vh93} predicts a slightly smaller amount of
limb darkening than the others. On the other hand,
the calculations of the I-band linear coefficients
are somewhat smaller than the measurement except for \citet{cl98b} with
$\log\,g=\,4.0$. (However, to be consistent with a higher surface gravity
while maintaining its color, the source star should be in the disk, which is
inconsistent with our inferred proper motion.) Since $\cv$ and $\ci$ are not
independent (in both the theories and in our measurement), it is more
reasonable to compare the I and V band measurements to the theories
simultaneously. Using the covariance matrix of the measurement of $\Gamma_{I}$
and $\Gamma_{V}$ (see \S 3), we derive error ellipses for our measurements
in the $(\ci,\,\cv)$ plane and plot them in Figure~10.
Formally, at the $1\,\sigma$ level, the calculations
of the linear limb-darkening coefficients in any of these models are not
consistent with our measurements. In principle, one could also constrain the
most likely stellar types that are consistent with the measured coefficients,
independent of \emph{a priori} information on the temperature and the gravity,
with a reference to a model. If we do this, the result suggests either that
the surface temperature is cooler than our previous estimate from the color
or that the source is a low-mass main-sequence ($\log\,g\geq\,4.0$) star.
However, the resulting constraints are not strong enough to place firm limits
on the stellar type even if we assume any of these models to be ``correct''.

One possible explanation of our general result -- the measured V-band
coefficients are nearly in perfect agreement with theories while the I-band
coefficients are only marginally consistent -- is non-linearity of stellar
limb darkening. Many authors have pointed out the inadequacy of the linear
limb darkening in producing a reasonably high-accuracy approximation of the
real stellar surface brightness profile \citep{wa85, di92, vh93, cl98b}.
Indeed, \citet{al99a} measured the two-coefficient square-root limb darkening
for a cusp-crossing microlensing event and found that the single-coefficient
model gives a marginally poorer fit to the data.
The quality of the linear parameterization has been investigated
for most theoretical limb-darkening calculations,
and the results seem to support this explanation.
\citet{vh93} defined the quality factors (Q in his paper)
for his calculations of limb-darkening coefficients,
and for 4000 K $\leq\Teff\leq$ 5000 K and 3.0 $\leq\log\,g\leq$ 4.0,
his results indicate that the linear parameterization is
a better approximation for V band than for I band.
Similarly, \citet{cl98b} provided plots of
summed residuals ($\sigma$ in his paper)
for his fits used to derive limb-darkening coefficients showing
that the V-band linear limb-darkening has lower $\sigma$ than I-band and
is as good as the V-band square-root limb-darkening near the temperature range 
of our estimate for the source of OGLE-1999-BUL-23.
In fact, \citet{di95} reported that the V-band limb darkening is
closest to the linear law in the temperature range $\Teff=\,4500\sim 4750$ K.
In summary, the source happens to be
very close to the temperature at which the linear limb darkening is
a very good approximation in $V$, but is less good in $I$.

The actual value of the
coefficient in the linear parameterization of a non-linear profile may vary
depending on the method of calculation and sampling. In order to determine
the linear coefficients, models (A) and (C) used a least square fit to
the theoretical (non-parametric) profile by sampling uniformly over 
$\cos\vartheta$ (see eq.~[3]), while model (B) utilized the principle of
total flux conservation between parametric and non-parametric profiles.
On the other hand, a fold-caustic crossing event samples the stellar surface
brightness by convolving it with a rather complicated magnification pattern
\citep{ga99}. Therefore, it is very likely that neither of the above samplings
and calculations is entirely suitable for the representation of
the limb-darkening measurement by microlensing unless the real intensity
profile of the star is actually same as the assumed parametric form (the
linear parameterization, in this case). In fact, the most apropriate way
to compare the measurement to the stellar atmospheric models would be
a direct fit to the (non-parametric) theoretical profile after
convolution with the magnification patterns near the caustics.
In the present paper, this has not been done, but we hope to make
such a direct comparison in the future.

\begin{acknowledgements}
We thank A. Udalski for re-reducing the OGLE data on very short notice
after we noticed an apparent discrepancy between the PLANET data and
the original OGLE reductions.
This work was supported by grants AST 97-27520 and AST 95-30619 from the NSF, 
by grant NAG5-7589 from NASA, by a grant from the Dutch ASTRON foundation
through ASTRON 781.76.018, by a Marie Curie Fellowship from the
European Union, and by ``coup de pouce 1999'' award from the Minist\`{e}re de
l'\'{E}ducation nationale, de la Recherche et de la Technologie.  
\end{acknowledgements}


\clearpage

\newpage
\begin{deluxetable}{ccrccrc}
\tablecaption{PLANET photometry of OGLE-1999-BUL-23}
\tablehead{
\colhead{telescope}&
\colhead{filter}&
\colhead{\# points}&
\colhead{normalization\,\tablenotemark{a}}&
\colhead{$b_{m}$\,\tablenotemark{b}}&
\colhead{$\hat{\eta}$\,\tablenotemark{c}}&
\colhead{$\theta_{m}$\,\tablenotemark{d}}\\
&&&&
\colhead{(\%)}&
\colhead{(arcsec$^{-1}$)}&
\colhead{(arcsec)}
}
\startdata
   SAAO&$I$&106&1.55&66.65& 0.0389&1.702\\
       &$V$& 47&2.27&98.65& 0.2136&1.856\\
  Perth&$I$& 39&1.00&81.26&      0&2.080\\
Canopus&$I$& 99&1.92&59.93&-0.1376&2.527\\
       &$V$& 35&1.88&92.06&-0.7626&2.587\\
   CTIO&$I$&103&1.44&41.14& 0.1794&1.715
\enddata
\tablenotetext{a}
{$\sigma_{{\rm normalized}}={\rm (normalization)}\times\sigma_{{\rm DoPHOT}}$}
\tablenotetext{b}
{Blending fraction at median seeing,
$b_{m}\equiv(F_{B,0}+\eta\,\theta_{m})/F_{S}$}
\tablenotetext{c}
{Scaled seeing correction coefficient,
$\hat{\eta}\equiv\eta/F_{S}$}
\tablenotetext{d}
{Median seeing disk size in FWHM}
\tablecomments
{The predicted flux of magnified source is
$F=F_{S}A+F_{B,0}+\eta\,\theta=
F_{S}\,[A+b_{m}+\hat{\eta}(\theta-\theta_{m})]$,
where A is magnification and $\theta$ is the FWHM
of the seeing disk in arcsec. The values of
$b_{m}$ and $\hat{\eta}$ are evaluated for the best model
(wide w/LD), in Table~2.}
\end{deluxetable}
\clearpage

\begin{deluxetable}{ccrcccccrrrc}
\rotate
\tablecaption{PLANET solutions for OGLE-1999-BUL-23}
\tablehead{
\colhead{$d$}&
\colhead{$q$}&
\colhead{$\alpha$\,\tablenotemark{a}}&
\colhead{$u_{0}$\,\tablenotemark{b}}&
\colhead{$\rho_{\ast}$}&
\colhead{$\te$}&
\colhead{$t_{0}$\,\tablenotemark{b}}&
\colhead{$t_{\cc}$}&
\colhead{$\Gamma_{I}$}&
\colhead{$\Gamma_{V}$}&
\colhead{$\Delta\chi^{2}$}&
\colhead{note}\\
&&\colhead{(deg)}&&\colhead{($\times 10^{-3}$)}&
\colhead{(days)}&\colhead{($\HJD'$)\,\tablenotemark{c}}&
\colhead{($\HJD'$)\,\tablenotemark{c}}&&&&
}
\startdata
2.42&0.39&74.63&0.90172&2.941&48.20&1356.154&1349.1062&
0.534&0.711&0.000&wide w/LD\,\tablenotemark{d}\\
0.56&0.56&260.35&0.10052&2.896&34.20&1344.818&1349.1063&
0.523&0.693&127.863&close w/LD\\
2.43&0.40&74.65&0.90987&2.783&48.48&1356.307&1349.1055&
0.&0.&172.815&wide no-LD
\enddata
\tablenotetext{a}
{The lens system is on the right-hand side of the moving source.}
\tablenotetext{b}
{the closest approach to the midpoint of the lens system}
\tablenotetext{c}
{$\HJD'\equiv\HJD-2450000.$}
\tablenotetext{d}
{LD $\equiv$ Limb Darkening}
\end{deluxetable}
\clearpage

\begin{deluxetable}{ccrcccccrrr}
\rotate
\tablecaption{Models in the neighborhood of the best-fit solution}
\tablehead{
\colhead{$d$}&
\colhead{$q$}&
\colhead{$\alpha$}&
\colhead{$u_{0}$}&
\colhead{$\rho_{\ast}$}&
\colhead{$\te$}&
\colhead{$t_{0}$}&
\colhead{$t_{\cc}$}&
\colhead{$\Gamma_{I}$}&
\colhead{$\Gamma_{V}$}&
\colhead{$\Delta\chi^{2}$}\\
&&\colhead{(deg)}&&\colhead{($\times 10^{-3}$)}&
\colhead{(days)}&\colhead{($\HJD'$)}&
\colhead{($\HJD'$)}&&&
}
\startdata
2.40&0.39&74.68&0.89020&2.998&47.41&1355.767&1349.1064&
0.560&0.730&0.616\\
2.41&0.38&74.59&0.89362&2.955&47.94&1356.010&1349.1062&
0.529&0.707&0.839\\
2.41&0.39&74.66&0.89587&2.968&47.80&1355.961&1349.1062&
0.533&0.709&0.481\\
2.41&0.40&74.72&0.89832&2.981&47.68&1355.909&1349.1065&
0.567&0.736&0.488\\
2.42&0.38&74.51&0.89912&2.930&48.37&1356.255&1349.1061&
0.522&0.701&0.474\\
\bf{2.42}&\bf{0.39}&74.63&0.90172&2.941&48.20&1356.154&1349.1062&
\bf{0.534}&\bf{0.711}&\bf{0.000}\\
2.42&0.40&74.68&0.90407&2.954&48.09&1356.111&1349.1065&
0.566&0.736&0.618\\
2.43&0.38&74.48&0.90489&2.904&48.78&1356.460&1349.1062&
0.523&0.702&0.612\\
2.43&0.39&74.62&0.90754&2.913&48.60&1356.335&1349.1061&
0.529&0.706&0.666\\
2.43&0.40&74.66&0.90974&2.923&48.48&1356.300&1349.1062&
0.532&0.709&0.916\\
2.44&0.39&74.54&0.91312&2.888&49.03&1356.580&1349.1062&
0.532&0.710&0.602
\enddata
\end{deluxetable}
\clearpage

\begin{figure}
\plotone{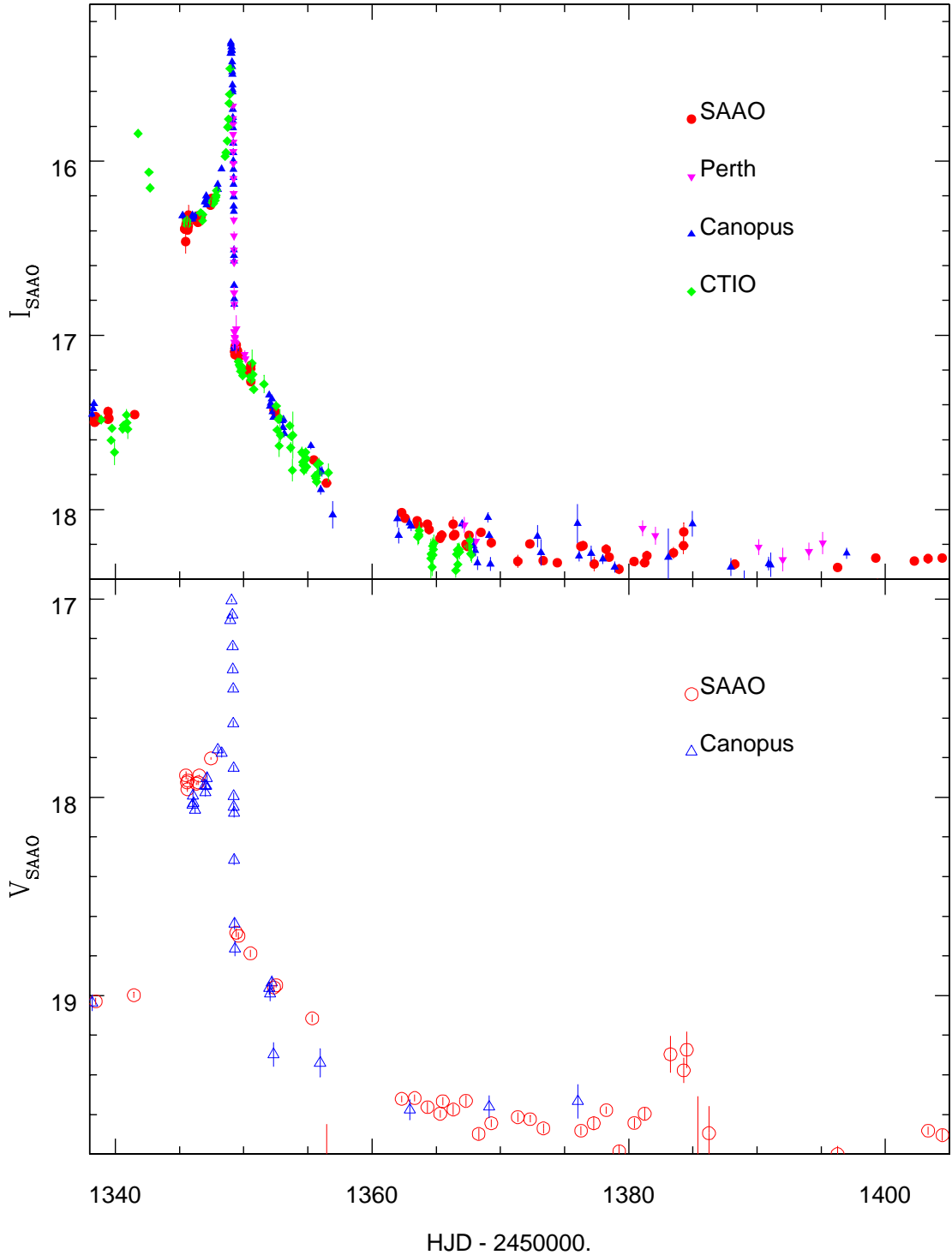}
\caption{
Whole data set excluding later-time baseline points ($1338<\HJD'<1405$),
in I (upper) and V (lower) bands.
Only the zero points of the different instrumental magnitude
systems have been aligned using the result of the caustic crossing fit
(\S 2.2); no attempt has been made to account for either different amounts
of blended light or seeing corrections.}
\end{figure}\clearpage

\begin{figure}
\plotone{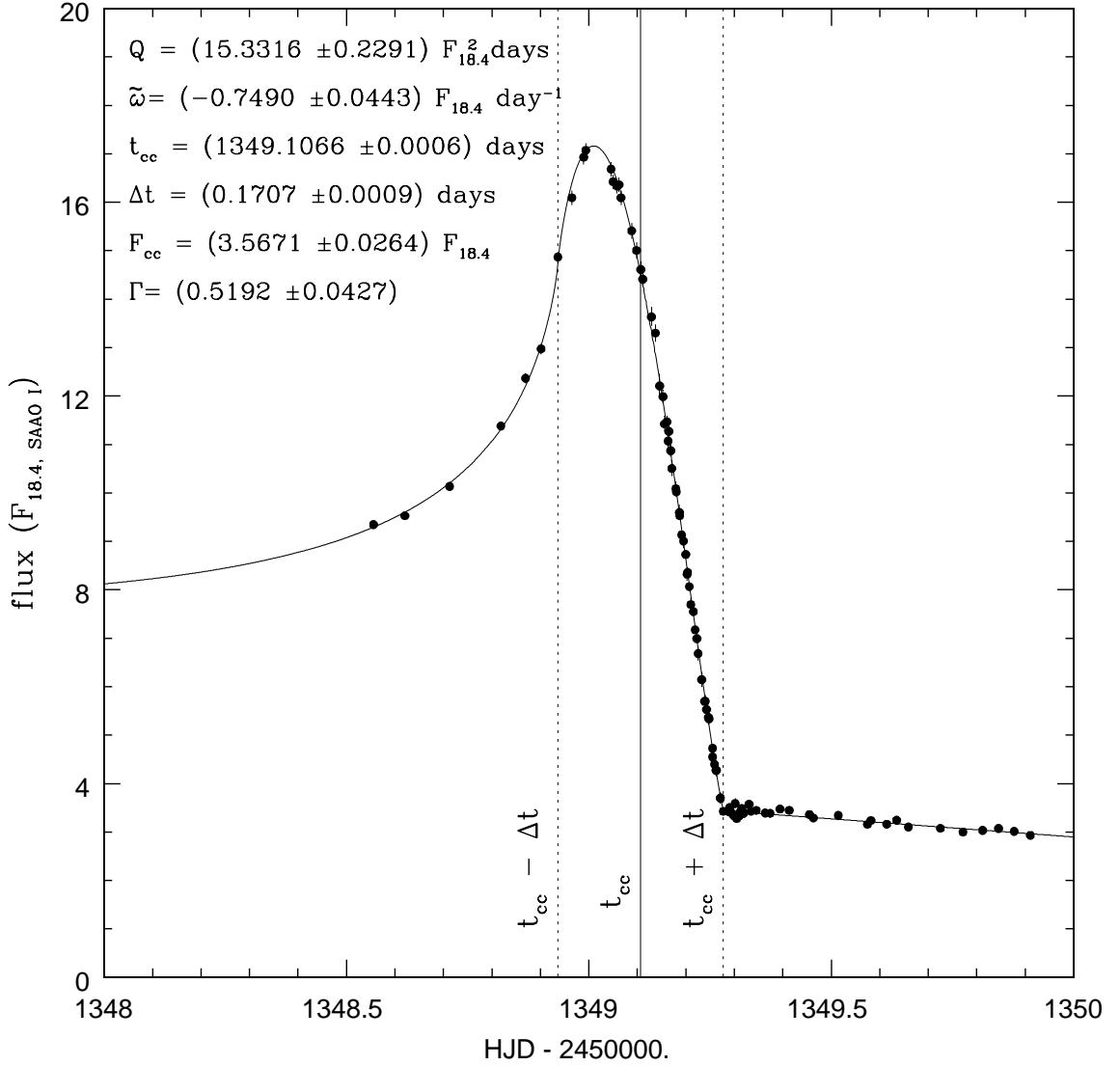}
\caption{
Fit of the caustic-crossing data to the six-parameter analytic
curve given by eq.~(1).
The time of second caustic crossing ($t_{\cc}$)
and the time scale of caustic crossing ($\Delta t$) are indicated by
vertical lines. The instrumental SAAO I-band flux, $F_{18.4}$, is given
in units of the zero point $I=18.4$.}
\end{figure}\clearpage

\begin{figure}
\plotone{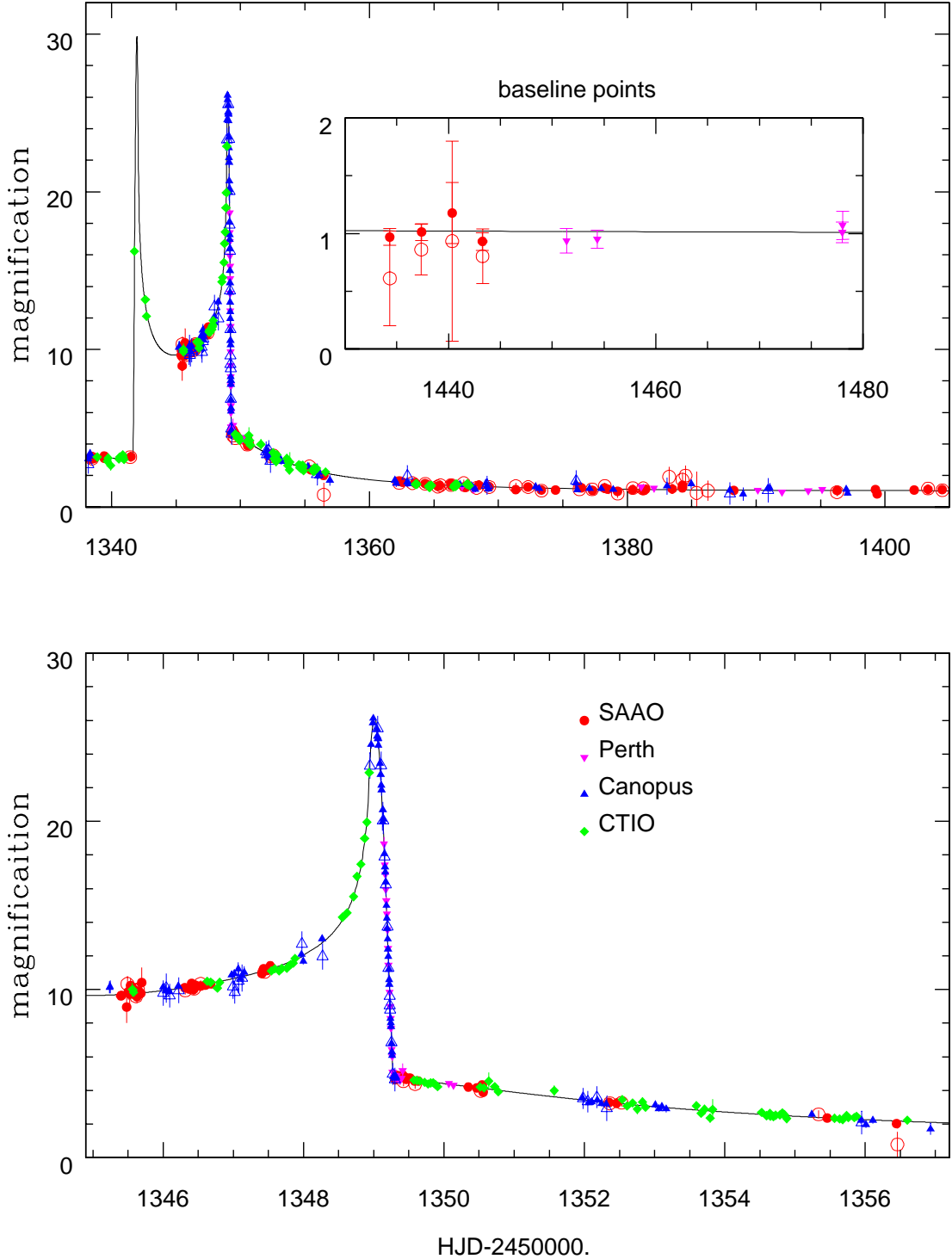}
\caption{
Magnification curve of the best-fit model taken from Table~2
for $(d,q)=(2.42,0.39)$.
Data included are SAAO (\emph{circles}), Perth (\emph{inverted triangles}),
Canopus (\emph{triangles}), and CTIO (\emph{diamonds}). Closed symbols are
for I band, and open symbols are for V band. Lower panel is a close-up of 
the time interval surrounding the second caustic crossing.}
\end{figure}\clearpage

\begin{figure}
\plotone{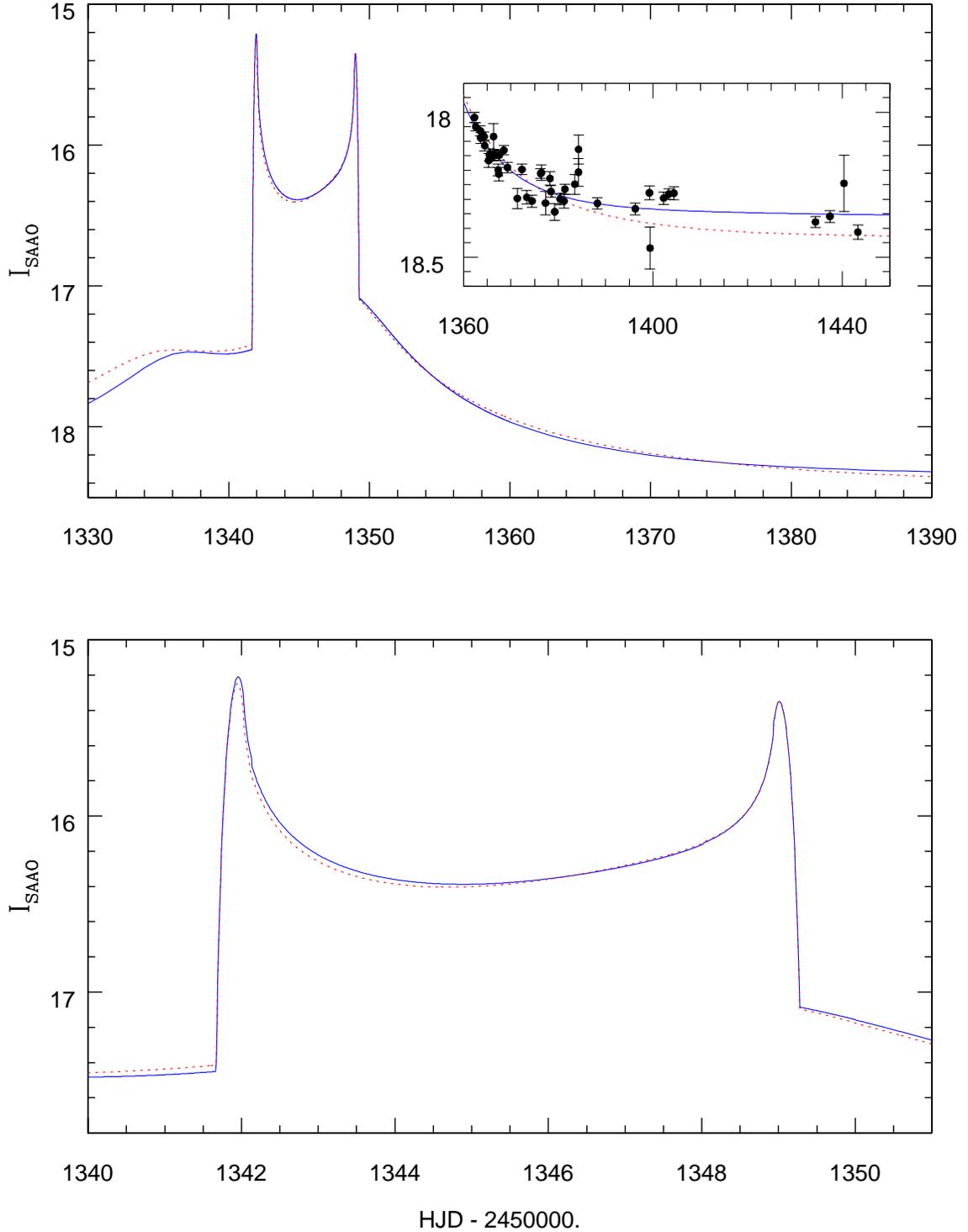}
\caption{
I-band light curves of the two ``degenerate'' models in SAAO instrumental
I magnitude. The solid line is the best-fit model of a wide binary lens,
$(d,q)=(2.42,0.39)$, and the dotted line is the close binary-lens model,
$(d,q)=(0.56,0.56)$. The filled circles are SAAO data points.
Both models are taken from Table~2 and use the estimates of blending
factors and baselines. The upper panel is for the whole light curve covered
by the data, and the lower panel is for the caustic-crossing part only.}
\end{figure}\clearpage

\begin{figure}
\plotone{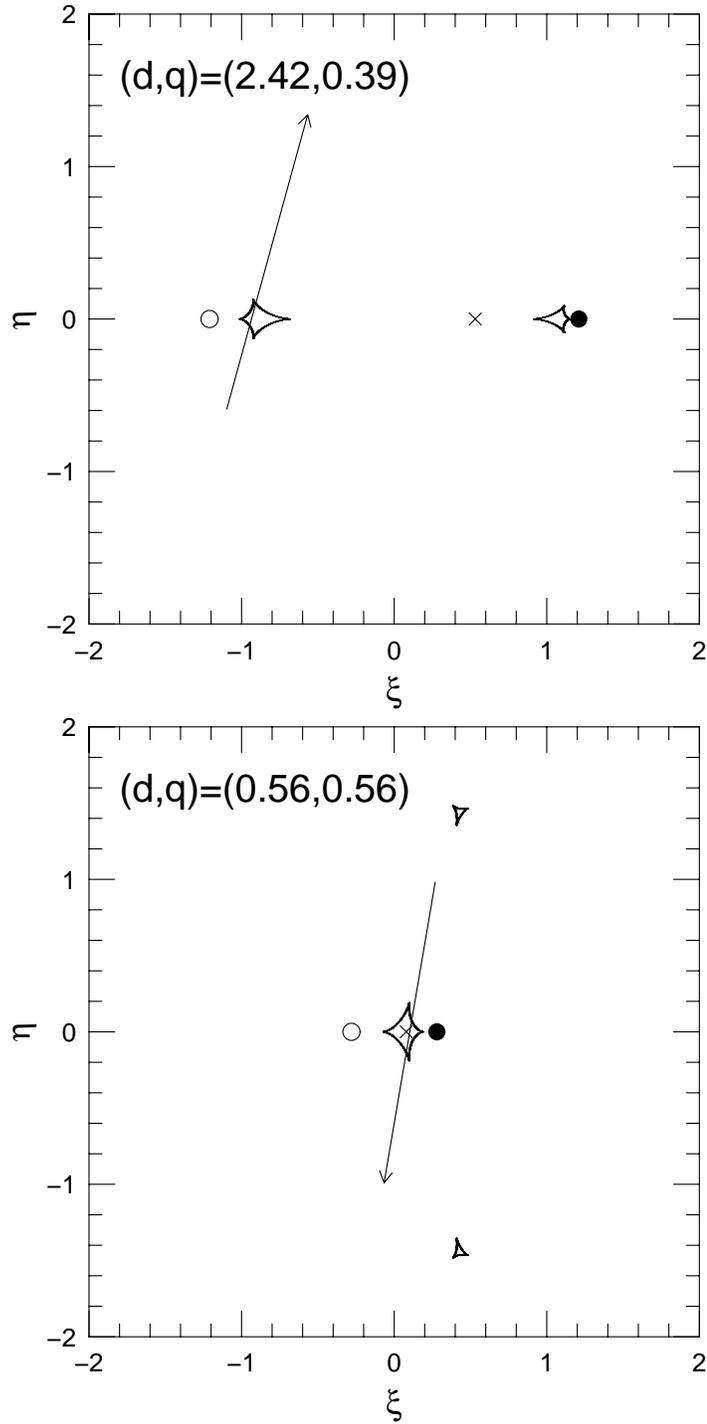}
\caption{
Lens geometries of the two ``degenerate'' models.
The origin of the coordinate system is the geometric center of the binary
lens, and `$\times$' is the center of the mass of the lens system.
One unit of length corresponds to $\thetae$. Closed curves are the caustics, 
the positions of the binary lens components are represented by circles,
with the filled circle being the more massive component. The trajectory
of the source relative to the lens system is shown by arrows, of which the
lengths are $2\thetae$.}
\end{figure}\clearpage

\begin{figure}
\plotone{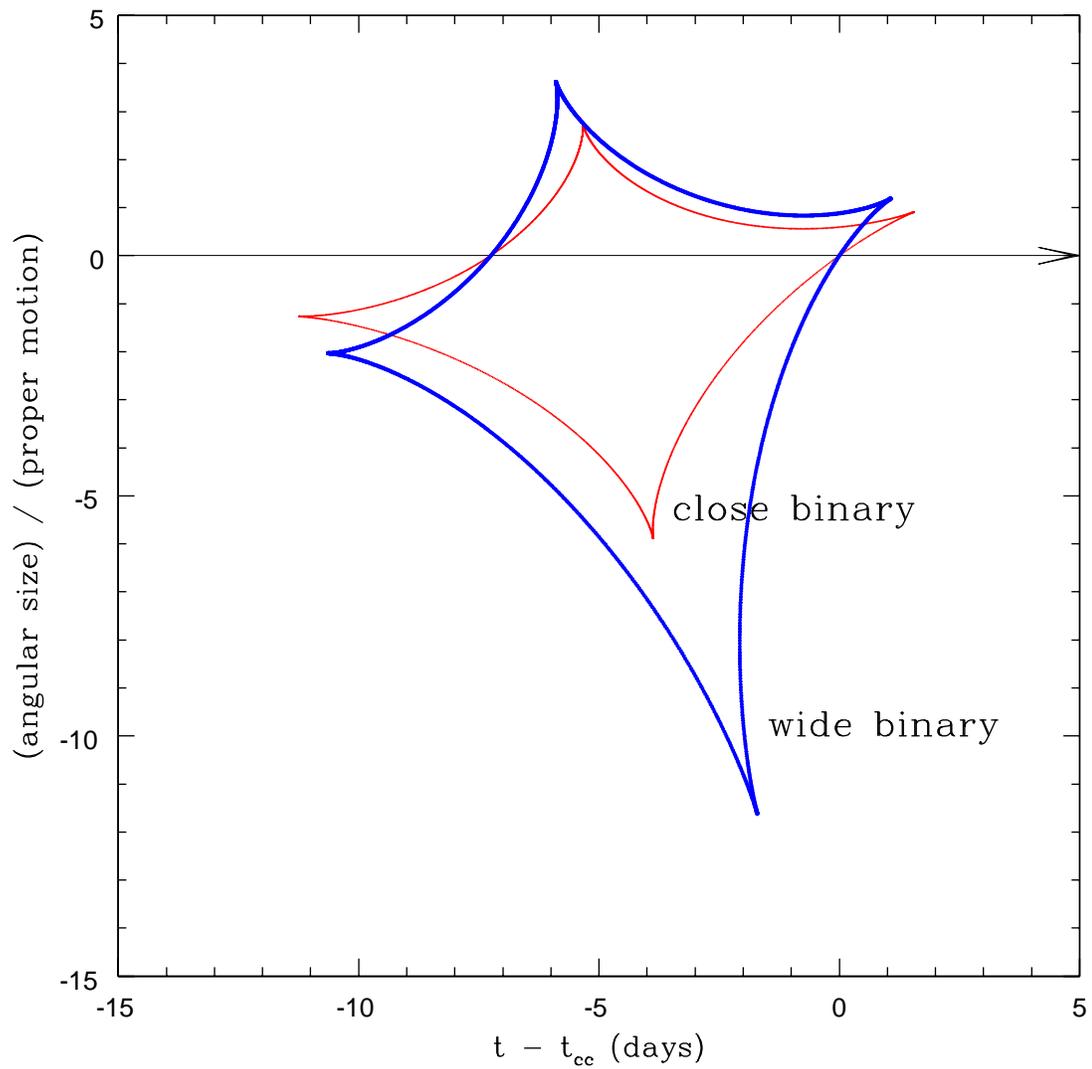}
\caption{
Caustics of the two ``degenerate'' models with respect to the source
path shown as a horizontal line. The similarity of the light curves
seen in Fig.~4 is due to the similar morphology of the caustics shown here.}
\end{figure}\clearpage

\begin{figure}
\plotone{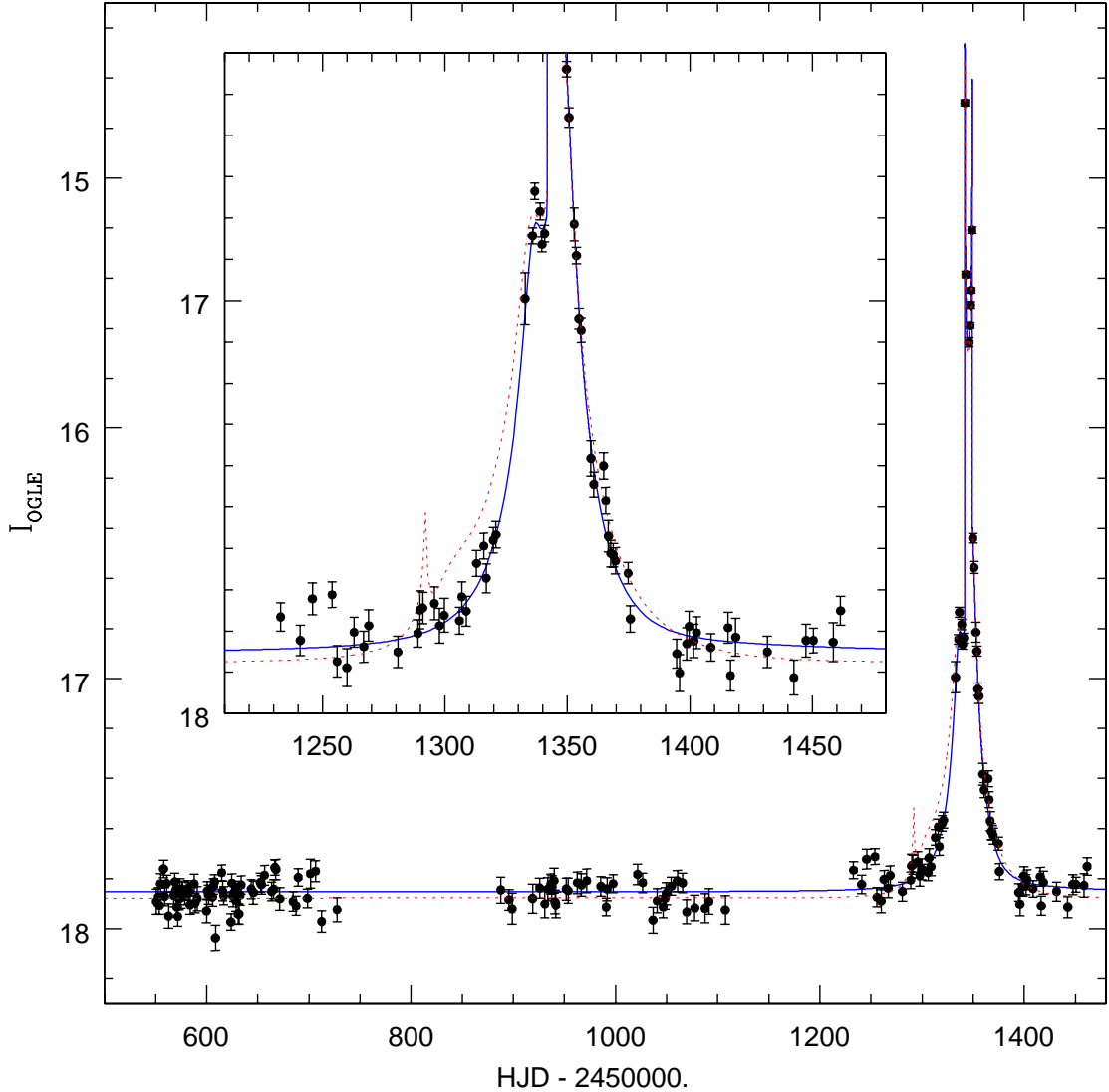}
\caption{OGLE observations of OGLE-1999-BUL-23.
The solid line is the light curve of the best-fit PLANET model (wide w/LD),
and the dotted line is the PLANET close binary model (close w/LD).
The models are determined by fitting PLANET data only, but the agreement
between the PLANET model (wide w/LD) and OGLE data is quite good. On the
other hand, OGLE data discriminate between the two ``degenerate''
PLANET models so that the wide-binary model is very much favored,
in particular, by the observations in $(1250<\HJD'<1330)$. The baseline of
the PLANET model (wide w/LD) is, $I_{OGLE}=17.852\pm 0.003$, which is
consistent with the value reported by OGLE, $I_{OGLE}=17.850\pm 0.024$.}
\end{figure}\clearpage

\begin{figure}
\plotone{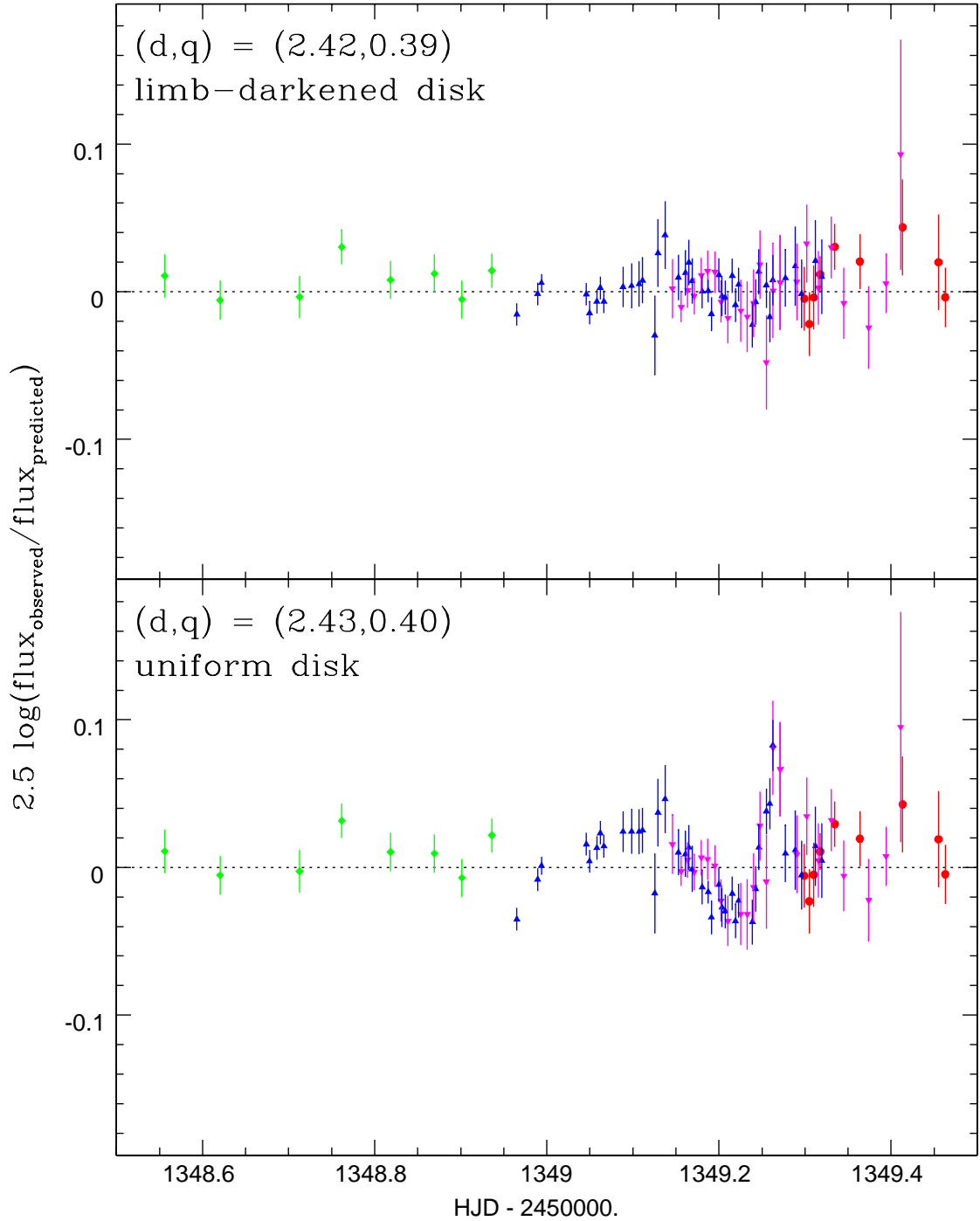}
\caption{
Residuals from PLANET models of OGLE-1999-BUL-23 around the second caustic
crossing. Upper panel shows the residual for a model incorporating
linear limb darkening (wide w/LD) and lower panel shows the same for
a uniform disk model (wide no-LD). Both models are taken from Table~2.
Symbols are the same as in Fig.~3. The residuals from the uniform disk are
consistent with the prediction that the source is limb-darkened while the
remaining departures from the limb-darkened model -- which are marginally
significant -- may be due to non-linearity in the surface brightness profile
of the source star.}
\end{figure}\clearpage

\begin{figure}
\plotone{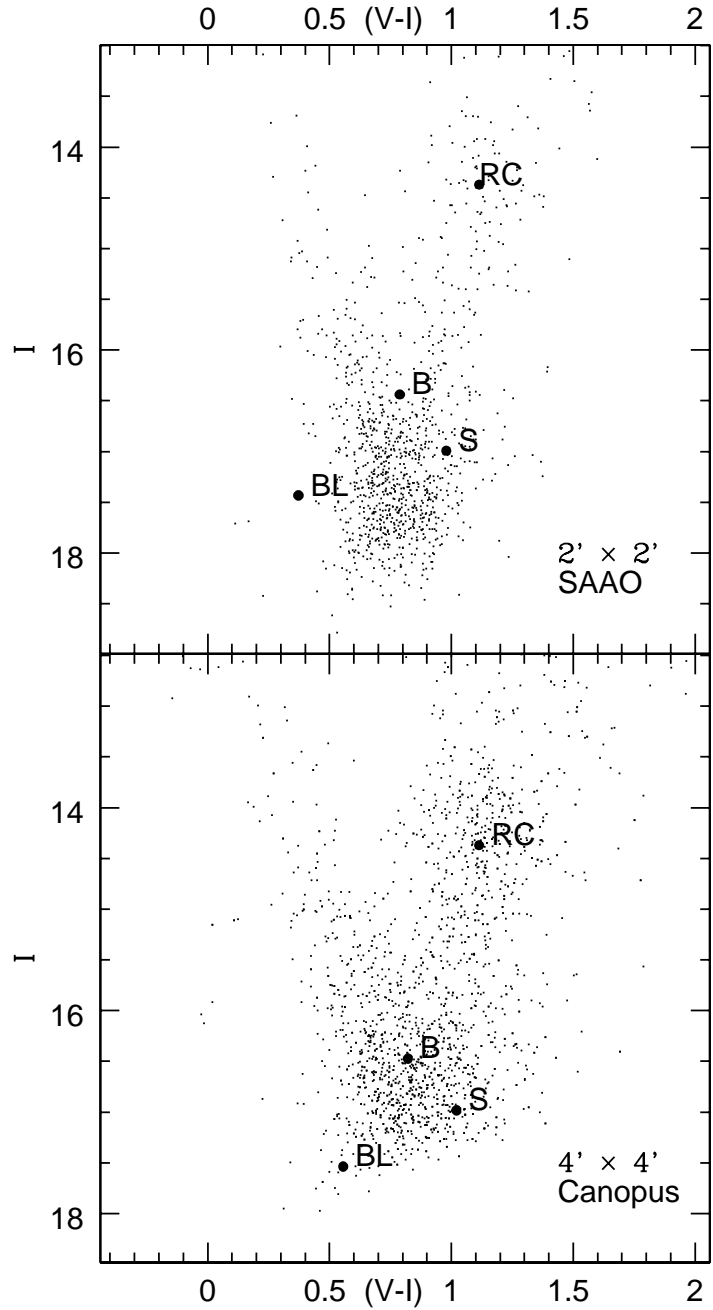}
\caption{
Color-magnitude diagram (CMD) of the field centered on OGLE-1999-BUL-23.
Upper CMD is derived from $2'\times 2'$ SAAO images, and lower CMD is from
$4'\times 4'$ Canopus images. The positions of the unlensed source (S),
the baseline (B), blended light (BL) at median seeing, and the center of
red clump giants (RC), are also shown. The extinction inferred from the
(reddened) OGLE magnitude of the source in the I band is, $A_{I}=1.18$,
which implies $E(V-I)=0.792$ assuming the extinction law,
$A_{I}=1.49E(V-I)$.}
\end{figure}\clearpage

\begin{figure}
\plotone{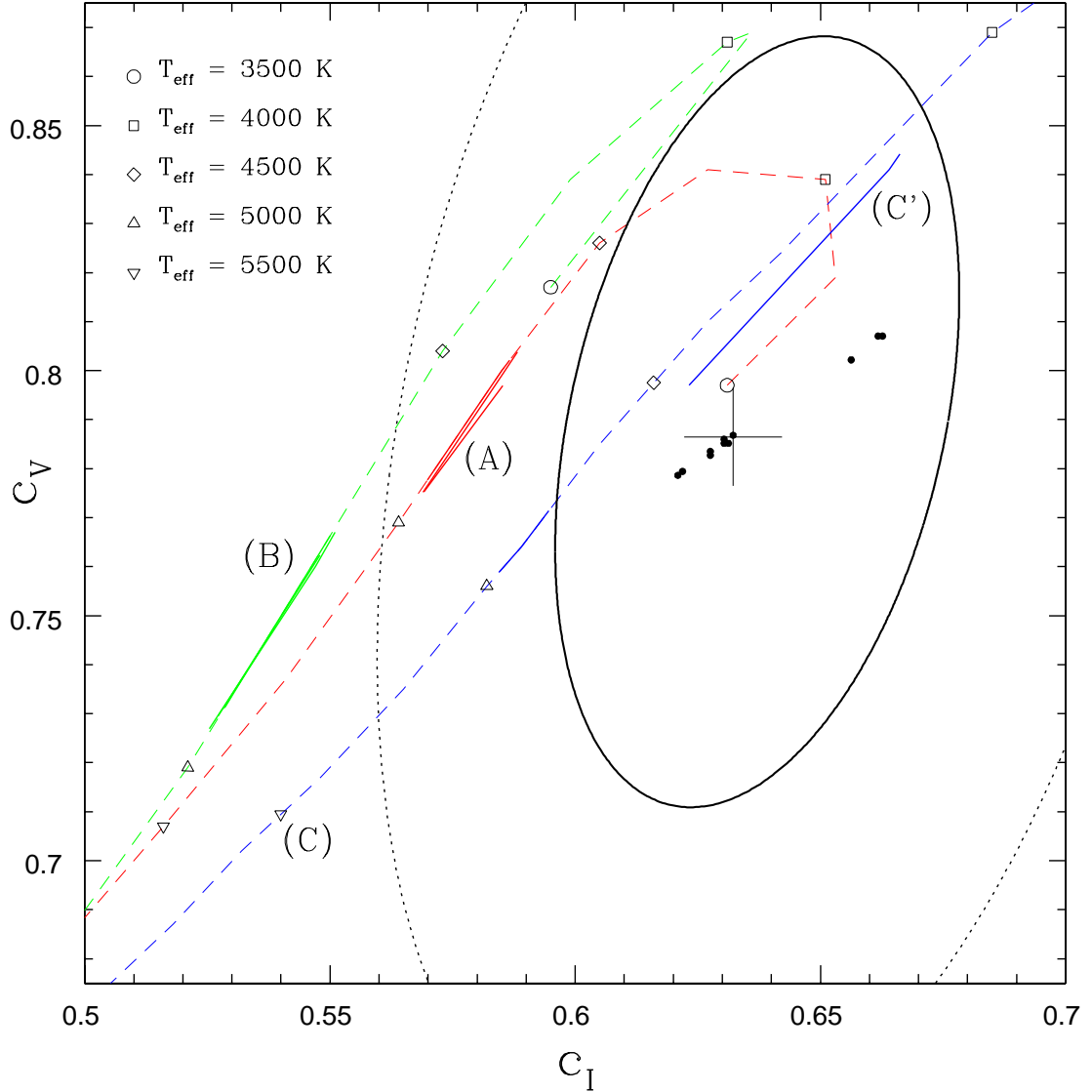}
\caption{
Comparison of linear limb-darkening coefficients. The measured value from
the best model is represented by a small cross. One (\emph{solid line})
and two (\emph{dotted line}) $\sigma$ error ellipses
are also shown. Small dots are the results with
different global parameters taken from Table~3. Various model predictions
are displayed by dashed lines ($\log\,g=\,3.5$). Model (A) is taken from
\citet{di95} and \citet{cl95}, (B) is from \citet{vh93}, and (C) is from
\citet{cl98b}. In particular, the predicted values in the temperature range
that is consistent with our color measurements ($\Teff=\,[4820\pm 110]$ K for
$\log\,g=\,3.0$; $\Teff=\,[4830\pm 100]$ K for $\log\,g=\,3.5$; and
$\Teff=\,[4850\pm 100]$ K for $\log\,g=\,4.0$) are emphasized by thick
solid lines. Model (C') is by \citet{cl98b} for stars of
$\Teff=\,(4850\pm 100)$ K for $\log\,g=\,4.0$. Although the measured value
of the limb-darkening coefficients alone favors this model, the model
is inconsistent with our estimation of the proper motion.}
\end{figure}\clearpage

\end{document}